\documentclass[10pt, conference]{IEEEtran}

\AtBeginDocument{%
  \providecommand\BibTeX{{%
    Bib\TeX}}}

\usepackage[ruled,linesnumbered]{algorithm2e}
\usepackage{multirow}
\usepackage{listings}
\usepackage{booktabs}
\usepackage{multicol}
\usepackage{tabularx}
\usepackage{adjustbox}
\usepackage{subfig}
\usepackage{subcaption}
\usepackage{makecell}
\usepackage{enumitem}
\usepackage{calligra}
\usepackage{mdframed}
\usepackage{fontawesome}
\usepackage{amsmath}
\usepackage{amsfonts}
\usepackage{graphicx}
\usepackage{textcomp}
\usepackage{xcolor}
\usepackage{balance}
\usepackage{caption}
\usepackage{wrapfig}
\usepackage{colortbl}
\usepackage{forest}
\usepackage{array}
\usepackage[most]{tcolorbox}
\usepackage{pifont}
\usepackage{bigstrut}
\usepackage{newtxtext}
\usepackage{url}

\def\BibTeX{{\rm B\kern-.05em{\sc i\kern-.025em b}\kern-.08em
    T\kern-.1667em\lower.7ex\hbox{E}\kern-.125emX}}

\title{Multi-Agent End-to-End Vulnerability Management for Mitigating Recurring Vulnerabilities}

\author{
    Zelong Zheng\textsuperscript{$\ast$}, 
    Jiayuan Zhou\textsuperscript{$\dagger$}, 
    Xing Hu\textsuperscript{$\ast\ddagger$}, 
    Yi Gao\textsuperscript{$\ast$}, 
    Shengyi Pan\textsuperscript{$\ast$}\\
    \textsuperscript{$\ast$}Zhejiang University, Hangzhou, China \\
    \textsuperscript{$\dagger$}Queen's University, Canada \\
    \{zelongzheng, xinghu, gaoyi01, shengyi.pan\}@zju.edu.cn, 
    jiayuan.zhou@queensu.ca
}

\definecolor{lightgreen}{rgb}{0.9,1,0.9}
\definecolor{grayheader}{gray}{0.85}
\definecolor{codebackground}{RGB}{240,240,240} %

\newcommand{\appname}{\textsc{MAVM}\xspace}
\newcommand{\appnamec}{\textsc{MAVM-c}\xspace}
\newcommand{\appnamev}{\textsc{MAVM-v}\xspace}

\newcommand{\portingagent}{Porting Agent\xspace}
\newcommand{\analyzingagent}{Analyzing Agent\xspace}
\newcommand{\consistencyagent}{Consistency-Check Agent\xspace}
\newcommand{\fixingagent}{Fixing Agent\xspace}
\newcommand{\validationagent}{Validation Agent\xspace}

\begin{document}


\maketitle

\def\thefootnote{$\ddagger$}\footnotetext{Corresponding author}

\begin{abstract}

Software vulnerability management has become increasingly critical as modern systems scale in size and complexity. 
However, existing automated approaches remain insufficient. 
Traditional static analysis methods struggle to precisely capture contextual dependencies, especially when vulnerabilities span multiple functions or modules. 
Large language models (LLMs) often lack the ability to retrieve and exploit sufficient contextual information, resulting in incomplete reasoning and unreliable outcomes. 
Meanwhile, recurring vulnerabilities emerge repeatedly due to code reuse and shared logic, making historical vulnerability knowledge an indispensable foundation for effective vulnerability detection and repair. 
Nevertheless, prior approaches such as clone-based detection and patch porting, have not fully leveraged this knowledge.
To address these challenges, we present \appname, a multi-agent framework for end-to-end recurring vulnerability management. 
\appname integrates five components, including a vulnerability knowledge base, detection, confirmation, repair, and validation, into a unified multi-agent pipeline. 
We construct a knowledge base from publicly disclosed vulnerabilities, thereby addressing the underuse of historical knowledge in prior work and mitigating the lack of domain-specific expertise in LLMs. 
Furthermore, we design context-retrieval tools that allow agents to extract and reason over repository-level information, overcoming the contextual limitations of previous methods. 
Based on agents, \appname effectively simulates real-world security workflows. 
To evaluate the performance of \appname, we construct a dataset containing 78 real-world patch-porting cases (covering 114 function-level migrations). 
On this dataset, \appname successfully detects and repairs 51 real vulnerabilities, outperforming baselines by 31.9\%–45.2\% in repair accuracy, which demonstrates its effectiveness.

\end{abstract}





\section{Introduction}


Software vulnerabilities have become a persistent threat as modern systems grow in size and complexity. 
High-profile incidents continue to demonstrate that vulnerabilities not only compromise personal privacy but can also disrupt critical infrastructure and inflict severe economic losses~\cite{cost_of_sv}. 
To mitigate such risks, mainstream development teams adopt structured vulnerability management workflows consisting of detection, confirmation, repair, and validation~\cite{securityfirefox2023}. 
While systematic, these workflows demand substantial human effort across phases, making them costly, error-prone, and too slow to keep pace with today’s rapid release cycles~\cite{valenzuelatoledo2024hiddencostsautomationempirical}.


A key observation is that many vulnerabilities are not unique: recurring vulnerabilities often emerge across projects due to code reuse or shared logic~\cite{redebug,tan2024similar,pan2024automating}. 
This drives extensive research into automating the processes of recurring vulnerability detection and repair. 
For example, ReDeBug~\cite{redebug} identifies unpatched code clones in OS-scale repositories, FVF~\cite{tan2024similar} filters out similar-but-patched (SBP) cases during detection, and PPatHF~\cite{pan2024automating} automates patch porting at the function level. 
While promising, these advances remain fragmented, each focusing narrowly on a single phase and lacking integration into a cohesive end-to-end process.
More fundamentally, current approaches fail on two critical challenges: 

\textbf{Challenge 1: Contextual reasoning is missing}. Vulnerabilities often depend on trigger chains spanning multiple functions as well as related data structures. 
However, static analysis typically provides only a partial view of such dependencies, and LLMs without tool support cannot retrieve sufficient repository context. 
As a result, critical conditions for the existence of a vulnerability may be overlooked.  

\textbf{Challenge 2: Historical vulnerability knowledge is underutilized}. Although recurring vulnerabilities frequently share root causes, prior methods rely narrowly on raw code changes, overlooking richer insights such as root cause explanations, triggering conditions, and mitigation strategies. 
Without exploiting this logic-level knowledge, automation remains shallow and prone to failure when the target repository environment differs from the original.  
These challenges prevent existing techniques from achieving effective end-to-end automation, making human intervention still indispensable.

To overcome these challenges, we propose a novel \textbf{M}ulti-\textbf{A}gent framework for end-to-end Recurring \textbf{V}ulnerability \textbf{M}anagement (\textbf{\appname}), which leverages multiple LLM-based agents to collaboratively perform vulnerability management tasks, effectively replacing human efforts. 
To the best of our knowledge, this is the first work to explore end-to-end automation of the entire vulnerability management process. 
\appname mainly consists of five agents, a vulnerability detector, and a vulnerability knowledge base (VKB): 
\ding{182} \textbf{VKB} stores comprehensive information of historical vulnerabilities, solving the challenge of underutilized historical vulnerability knowledge. 
\ding{183} \textbf{Vulnerability Detector} performs a rapid screening of the target repository to identify potential recurring vulnerabilities based on historical vulnerabilities stored in the VKB.
\ding{184} \textbf{\portingagent} ports the analysis points from the historical vulnerability scene to the recurring vulnerability scene. 
\ding{185} \textbf{\analyzingagent} executes the new analysis points to further confirm whether the detection result is truly vulnerable. 
\ding{186} \textbf{\consistencyagent} checks whether the functions involved in the historical patch are consistent with those in the target repository.
\ding{187} \textbf{\fixingagent} fixes the vulnerability with the help of the analysis process, consistency-check report, and historical information. 
\ding{188} \textbf{\validationagent} checks whether the repair is correct and returns feedback if not.
With the support of tools designed to retrieve contextual information from repositories, the agents can obtain the necessary context, addressing the issue of missing contextual reasoning in vulnerability management. 
These designs simulate real-world workflows and overcome the above challenges, enabling end-to-end vulnerability management.

To evaluate the effectiveness of \appname, we construct an evaluation dataset from real-world examples of patch migration. 
Specifically, we filter data based on the Mystique~\cite{mystique} dataset and collect additional datasets following the dataset construction method of Pan et al.~\cite{pan2024automating}. 
The final dataset involves 13 repositories, containing 78 patch migration examples at the commit level and 114 function-level vulnerable function pairs. 
Compared with the baselines (including GPT-4o~\cite{gpt4o}, PPatHF, Mystique, and FVF), in the final repair accuracy metric, \appname outperforms the baselines by 31.9\%–45.2\%, which demonstrates its effectiveness.
We summarize our contributions as follows:


\begin{itemize}[leftmargin=*]
  \item We propose \appname, the first multi-agent framework designed for end-to-end recurring vulnerability management. 
  By coordinating specialized agents, \appname effectively simulates real-world workflows. 
  Leveraging the VKB of real-world cases and context-retrieval tools that empower agents for repository-level reasoning, \appname effectively addresses the challenges of contextual insufficiency and underutilized historical vulnerability information.
  \item We construct a \textit{benchmark dataset} for evaluating end-to-end vulnerability management. The dataset covers 13 widely used open-source repositories with GitHub stars ranging from 562 to 91.8k. It includes 78 patch porting cases and 114 function-level vulnerability pairs, providing a rigorous and diverse basis for future research.
  \item The experimental results demonstrate that \appname outperforms state-of-the-art baselines by \textit{31.9\%–45.2\%} in the final repair accuracy metric, validating its effectiveness and practical potential.  
\end{itemize}

\section{Background and Motivation}
\label{sec:background}

\begin{figure*}
    \centering
  \includegraphics[width=1\textwidth]{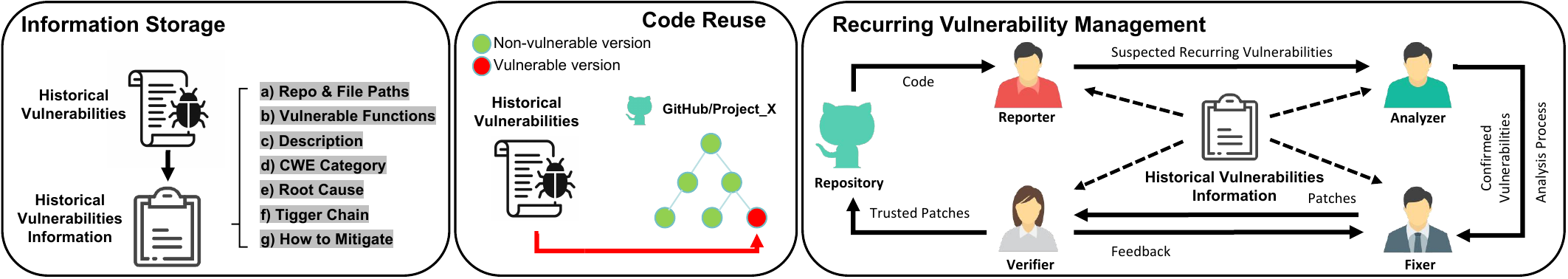}
  \caption{The background and basic process of recurring vulnerability management.}
  \label{fig:vulnmanage}
\vspace{-0.3cm}
\end{figure*}

This section introduces the background and the motivation of our work.

\subsection{Recurring Vulnerability Management}
\label{sec:recurringvuln}
\textbf{Recurring Vulnerabilities}. 
Recurring vulnerabilities refer to security flaws that share similar characteristics and reappear across different software projects or repositories. 
This phenomenon is primarily driven by the prevalent practice of code reuse and shared code logic in the open-source ecosystem ~\cite{redebug,tan2024similar}, such as the creation of hard forks~\cite{pan2024automating}. 
When a vulnerable code snippet is cloned, moving from a source project to a target repository, the original flaw is spread to new environments. 
This process creates a widespread security risk across various platforms.~\cite{vuddy}.

\textbf{Recurring Vulnerability Management}. 
Systematic recurring vulnerability management is essential throughout the software lifecycle, aiming to detect, confirm, and fix recurring software vulnerabilities. 
As shown in Figure~\ref{fig:vulnmanage}, it is an iterative pipeline where each phase is tightly connected. 
Effective management ensures that information from the original vulnerability is deeply analyzed and stored for subsequent recurring vulnerability management in target repositories. 
Meanwhile, diagnostic insights gained during the detection phase, such as triggering conditions and root causes of the vulnerability, can be transferred to guide the repair stage. 
This integrated approach effectively maintains the integrity of the software supply chain.

\subsection{Existing Works and Challenges}
The prevalence of cross-repository security risks drives extensive research into automating the processes of recurring vulnerability detection and repair. 
Historically, these tasks have been addressed by two separate research lines that function as independent point solutions~\cite{10.1145/1287624.1287634, redebug, mvp, 10.1145/3460319.3464821, 10.1145/3576915.3623188}, and a comprehensive end-to-end approach for the entire vulnerability management lifecycle remains absent.

\textbf{Recurring Vulnerability Detection}. 
Existing approaches generally identify recurring vulnerabilities by matching the source code of a target system with known vulnerabilities. 
These methods can be classified into two primary technical routes. 
The first is clone-based detection~\cite{redebug, 10.1145/1287624.1287634, mvp}, which treats recurring vulnerability detection as a code clone detection problem. 
The second route is function-matching-based detection~\cite{tan2024similar}, where techniques directly use vulnerable functions as signatures and detect threats by measuring the similarity score between the vulnerable function and the target function. 
While efficient, these tools often fail to differentiate the minute differences between vulnerable and patched functions, leading to high false positives.

\textbf{Recurring Vulnerability Repair (Patch Porting)}. 
Once a recurring flaw is located, the next objective is to port the patch to the target project automatically. 
Patch porting is essentially a process of adapting security fixes across divergent codebases. 
Methods like PPatHF~\cite{pan2024automating} and Mystique ~\cite{mystique} are both LLM-based method aimed at patch porting.
This task is particularly challenging in scenarios like hard forks, where an understanding of different implementations of the same functionality is required to ensure the patch remains functional in the new environment.

\textbf{Challenges}. 
Despite their specialized capabilities, there are still some challenges in achieving end-to-end vulnerability management. 
\ding{182} \textbf{contextual reasoning is often missing}. 
Since vulnerabilities frequently depend on complex trigger chains spanning multiple functions, a shallow view of code similarity is insufficient to confirm the existence of a recurring flaw or to adapt its fix.
This renders most existing approaches in the relevant works produce inaccurate detection or repair results. 
For example, ReDeBug~\cite{redebug} uses pure syntax-level matching to find recurring vulnerabilities, and it ignores essential context. 
Furthermore, the latest patch porting methods, PPatHF~\cite{pan2024automating} and Mystique~\cite{mystique}, are both intra-procedural. 
They focus on single-function-level analysis and cannot retrieve critical information residing outside the function boundary, such as global structure definitions.
\ding{183} \textbf{historical vulnerability knowledge is underutilized}. 
Prior methods rely narrowly on raw code snippets, overlooking richer logic-level insights such as root cause explanations and triggering conditions. 
By ignoring this, current automation remains prone to failure when the target repository environment contains structural variations compared to the source repository.
For instance, ReDeBug~\cite{redebug} merely treats historical patches as input for syntax-based similarity matching, leading to a high rate of false positives. 
Moreover, while PPatHF~\cite{pan2024automating} and Mystique~\cite{mystique} reference historical vulnerable and patched functions to guide repair, they often overlook the underlying security intent and deep information of the vulnerability. This neglect of deeper vulnerability knowledge frequently results in inaccurate or incomplete repairs.



\subsection{Motivation}


We observe unique opportunities in leveraging a multi-agent system and a vulnerability knowledge base to solve the aforementioned challenges.
\ding{182} Through the autonomous reasoning abilities of LLM-based agents and the use of specialized tools to retrieve global context, the challenge of insufficient information can be resolved. 
Agents can actively investigate function call chains and data structures beyond the specific function where the vulnerability resides, effectively emulating the comprehensive analysis performed by security experts.
\ding{183} By performing deep analysis of historical vulnerabilities and storing logic-level insights (such as root cause explanations) in a VKB, the system can perform recurring vulnerability management more effectively.

Furthermore, a single agent can assume responsibility for a specific phase of vulnerability management, and multiple agents can simulate security development teams manage vulnerabilities in practice.
Therefore, we propose MAVM, a novel multi-agent-based end-to-end framework for recurring vulnerability management.


\begin{figure*}
\centering
  \includegraphics[width=1\textwidth]{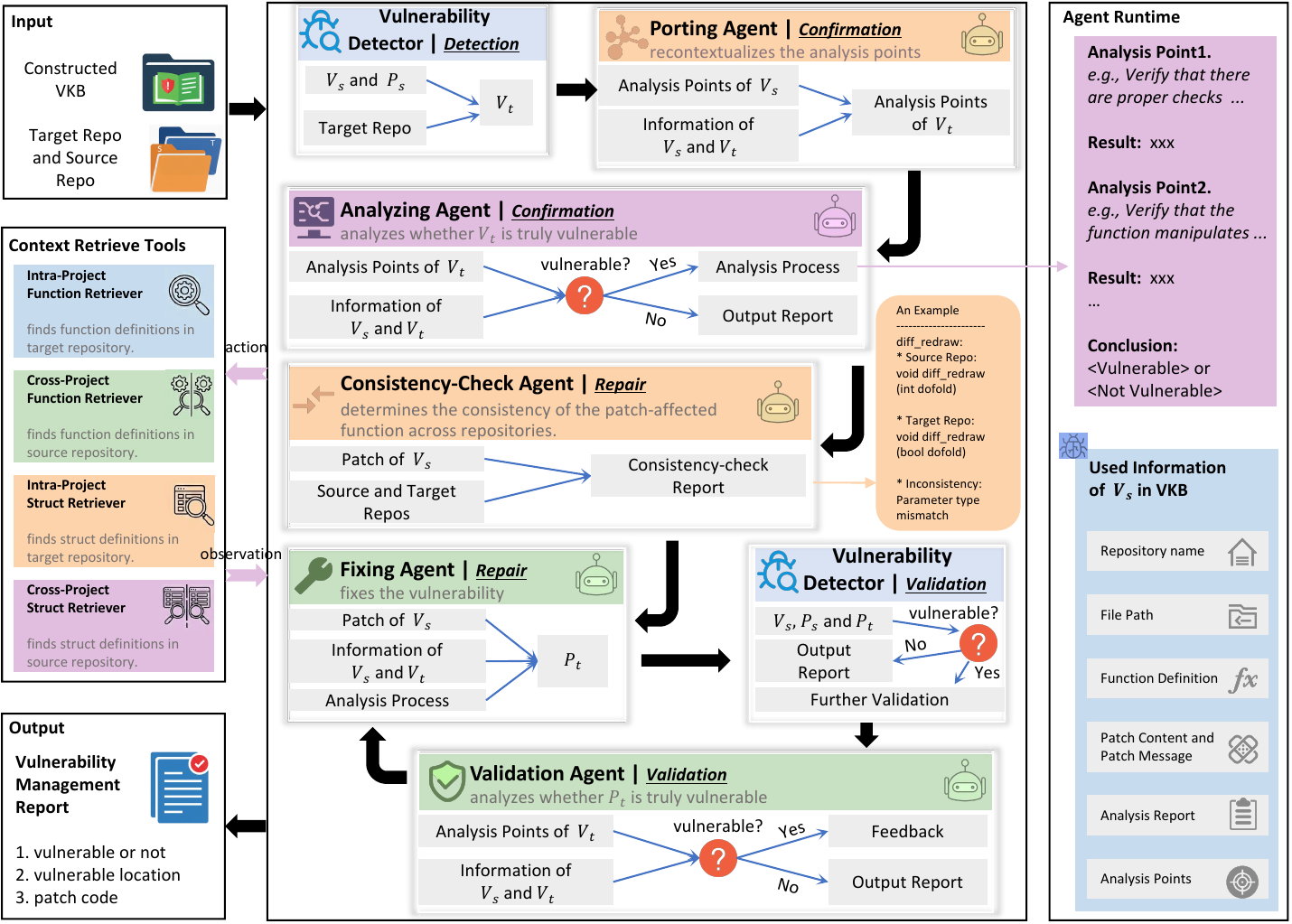}
  \caption{Overview of our approach.}
  \label{fig:method}
  \vspace{-0.4cm}
\end{figure*}

\section{Approach}
\label{sec:approach}


This section presents \appname, our multi-agent framework that automates end-to-end recurring vulnerability management.

\subsection{Overview}
\label{sec:overview}
\appname is designed to perform recurring vulnerability management on a target code repository based on publicly disclosed vulnerability information. 
\appname advances beyond traditional vulnerability scanning and patching by (1) fully leveraging historical vulnerability information through the VKB, (2) integrating context-aware tools for agents to perform repository-level reasoning, and (3) first achieving the end-to-end recurring vulnerability management.

As shown in Figure~\ref{fig:method} and Figure~\ref{fig:db}, the input to \appname includes the disclosed CVE information (CVE ID and Git Commit URL), the codebases of the repositories corresponding to these CVEs, and the codebase of the target repository. 
The output is a vulnerability management report for the current version of the target repository. 
Specifically, the report provides: (1) whether the current version contains any vulnerabilities; (2) the file paths and functions where vulnerabilities are located (if any); and (3) the patches that address these vulnerabilities (if any). 
As shown in Figure~\ref{fig:method}, \appname consists of five components:

\ding{172} \textbf{Vulnerability Knowledge Base} (Section~\ref{sec:vkb}). We systematically construct a Vulnerability Knowledge Base (VKB) that encodes both basic attributes and higher-level analytical insights extracted from historical vulnerabilities. 

\ding{173} \textbf{Vulnerability Detection} (Section~\ref{sec:detectandconfirm}). 
\appname uses ReDeBug~\cite{redebug} and the hash-based method from Tan et al.~\cite{tan2024similar} to rapidly detect recurring vulnerabilities in the target repository. 

\ding{174} \textbf{Vulnerability Confirmation} (Section~\ref{sec:detectandconfirm}). 
After identifying a potential vulnerability \( V_t \) in the target repository based on a known disclosed vulnerability \( V_s \) from the VKB, the \portingagent first transfers the analysis points of \( V_s \) to the context of \( V_t \) within the target repository. 
This enables the generation of new analysis points specific to \( V_t \). 
Then, the \analyzingagent determines whether \( V_t \) constitutes a real vulnerability by leveraging these new analysis points along with additional information of \( V_s \) from the VKB.

\ding{175} \textbf{Vulnerability Repair} (Section~\ref{sec:repairandvalidation}). 
Once \( V_t \) has been confirmed as a vulnerability, we obtain the vulnerable function \( VF_t \). 
At the same time, the corresponding patch for \( V_s \) can be retrieved from the VKB. 
Then a function-level consistency check is performed between the repositories of \( V_s \) and \( V_t \) to determine whether the functions involved in the patch differ across the two codebases. 
Based on the analysis process, the consistency-check report, and historical patch, the \fixingagent generates a patch to repair \( VF_t \).

\ding{176} \textbf{Patch Validation} (Section~\ref{sec:repairandvalidation}). Finally, the generated patch is passed through the vulnerability detector and the \validationagent to verify whether the vulnerability has been eliminated. 
If the check passes, the process concludes with the generation of a vulnerability management report summarizing the entire workflow. 
If any issues are found, they are fed back to the \fixingagent for re-patching.

\subsection{Vulnerability Knowledge Base}
\label{sec:vkb}

\begin{figure}
  \includegraphics[width=\linewidth]{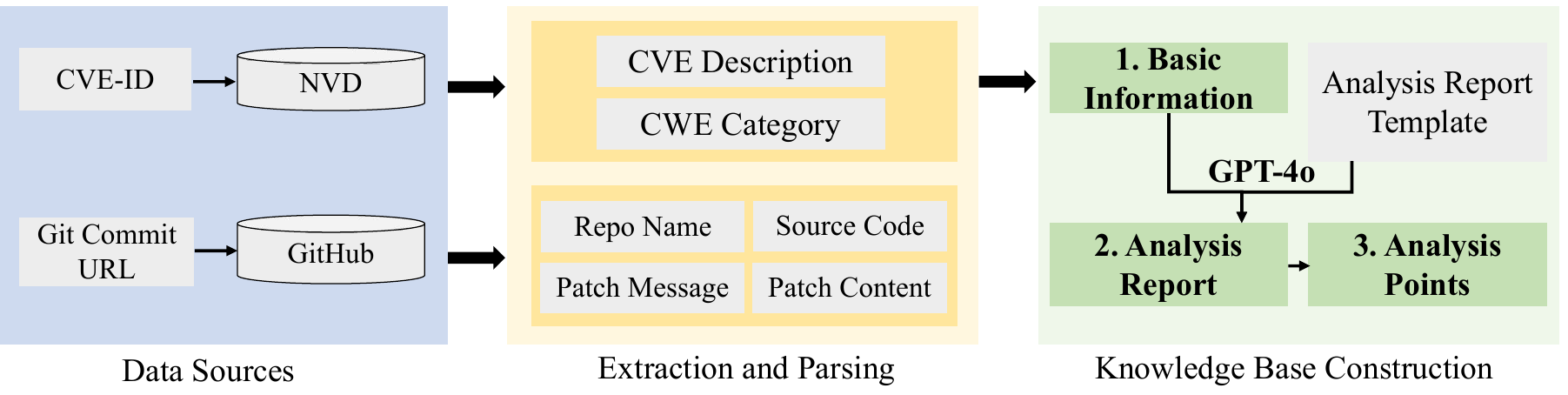}
  \caption{Vulnerability knowledge base construction.}
  \label{fig:db}
  \vspace{-0.5cm}
\end{figure}

\begin{figure*}
\centering
  \includegraphics[width=1\linewidth]{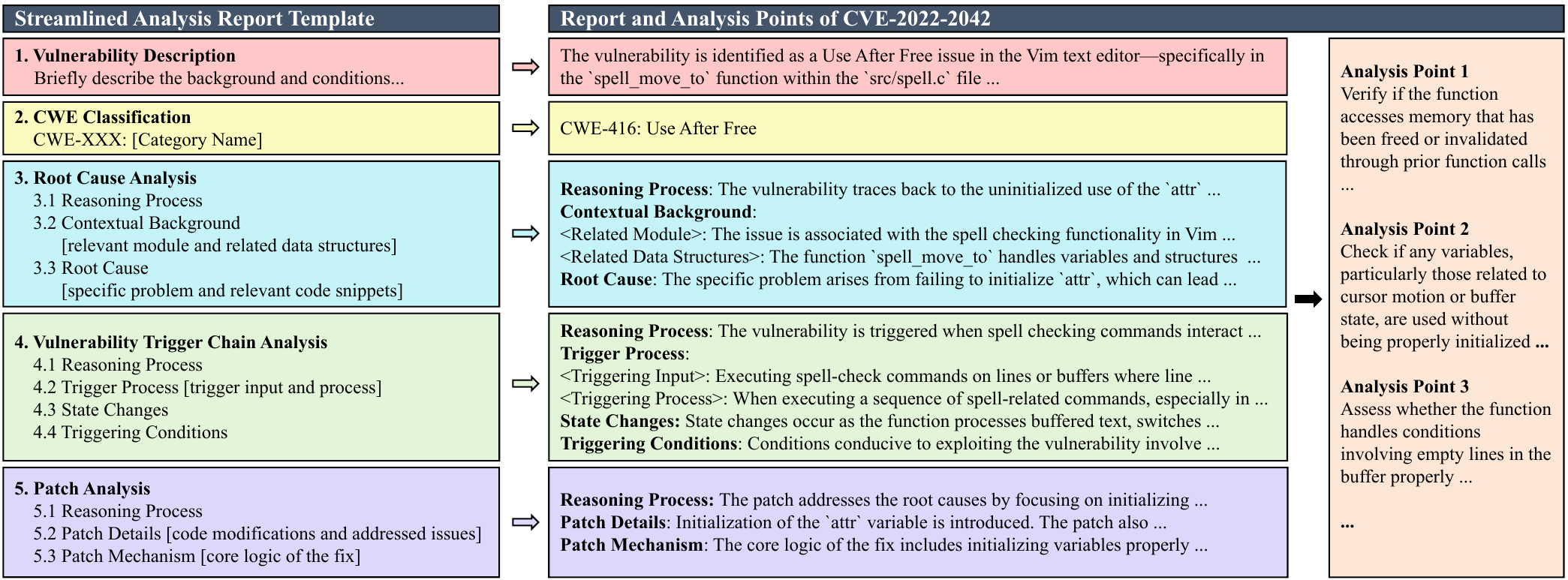}
  \caption{The analysis report template designed for VKB, the report and analysis points generated for CVE-2022-2042 by GPT-4o.}
  \label{fig:template}
    \vspace{-0.3cm}
\end{figure*}

Since our focus is on recurring vulnerabilities, fully acquiring historical vulnerability information is essential, as a thorough understanding of historical vulnerabilities is critical to determining whether the current codebase contains recurring ones. 
This task is non-trivial as it involves addressing two key challenges: (1) determining the types of information that need to be collected, and (2) determining the methods for acquiring such information.

To achieve a comprehensive collection of historical vulnerability information and to tackle these challenges, the VKB is designed and constructed. 
Regarding the first challenge, after extensively reviewing publicly available vulnerability analysis reports on the web, we classify the required information into two categories. 
The first is basic information, which refers to the intrinsic attributes of a vulnerability, such as the vulnerable location and the vulnerable code. 
The second is information that can only be derived through further analysis, such as the root cause of the vulnerability. 

As illustrated in Figure~\ref{fig:db}, the initial inputs for VKB construction are the CVE ID and the Git Commit URL. 
Using the CVE ID, we retrieve the CVE description and its associated CWE category from the NVD~\cite{nvd}. 
The Git Commit URL allows us to obtain the repository name, file path, vulnerable function code, the patch content, and the corresponding commit message. 
These elements collectively constitute the basic information in the VKB. 
Based on this basic information and a predefined multi-perspective vulnerability analysis template (described later in this section), we employ GPT-4o~\cite{gpt4o} to generate vulnerability analysis reports automatically. 
Subsequently, we leverage the same LLM to automatically distill these reports into concise analysis points (also detailed later in this section). 
GPT-4o is specifically chosen for its superior zero-shot reasoning capabilities and high efficiency in processing complex, structured analysis requests~\cite{gpt4o}. 
Finally, the VKB consists of three components: the basic information, the vulnerability analysis report, and the analysis points.

\textbf{Vulnerability Analysis Reports}. 
Since our work focuses on recurring vulnerabilities, thoroughly analyzing historical vulnerabilities is crucial. 
However, public vulnerability reports often lack depth and provide limited insights for understanding the vulnerabilities. 
For example, the report of CVE-2015-8718 on the National Vulnerability Database (NVD) contains only a brief description, the CWE category, and a CVSS score, and mostly uninformative reference links, with only the Git link offering patch information but no in-depth analysis~\cite{cve-2015-8718}. 
To address this, we design a unified multi-perspective report template capturing both intrinsic and causal properties of vulnerabilities. 
We establish the structured schema through a systematic manual analysis of 100 representative CVEs. 
To ensure the reliability and practical relevance of this schema, we adopt a data-driven inductive approach. 
Specifically, we source the dataset of these 100 CVEs by selecting those accompanied by high-quality, expert-authored analysis reports from technical security communities and blogs (e.g., ~\cite{cvereport1, cvereport2, cvereport3}). 
By grounding our analysis in these detailed reports, we iteratively refine the analytical dimensions to align with real-world security auditing practices, ultimately identifying three critical deep-analysis dimensions: \textbf{root cause analysis}, \textbf{vulnerability trigger chain analysis}, and \textbf{patch analysis}. 
Integrating these derived dimensions with the essential vulnerability context, we define the final multi-perspective template. 
As shown in Figure~\ref{fig:template}, this template consists of five key sections: the foundational vulnerability description and the CWE category, followed by the three dimensions derived from expert reports. 
This comprehensive structure enables the LLM to automatically generate rich, consistent analysis reports at scale from raw CVE and commit data (see Figure~\ref{fig:template} for an example), laying a solid foundation for subsequent vulnerability management tasks.

\textbf{Vulnerability Analysis Points}. 
Vulnerability Analysis Points serve as actionable diagnostic heuristics that guide the agents in determining the presence of a vulnerability. 
Instead of providing abstract summaries, these points define the specific logic states and critical execution conditions, such as API misuse patterns or boundary errors, which must be verified to characterize the vulnerability. 
By utilizing structured directives extracted via GPT-4o (see Figure~\ref{fig:template} for an example), the \portingagent can systematically map the diagnostic requirements of a historical vulnerability to the current context. 
This granularity ensures that the \analyzingagent follows a rigorous reasoning path to evaluate whether the target code replicates the fundamental flaw of the original vulnerability.



\subsection{Vulnerability Detection and Confirmation}
\label{sec:detectandconfirm}

\subsubsection{Vulnerability Detection}
\appname integrates ReDeBug~\cite{redebug} and the hash-based method proposed by Tan et al.~\cite{tan2024similar} as our core detector because they are both practical recurring vulnerability detection tools and can increase coverage by lowering the retrieval threshold parameter to maximize coverage.
Redebug can quickly find unpatched vulnerability code clones in code bases by using syntax-based pattern matching~\cite{redebug}. 
The hash-based method can rapidly identify suspected vulnerable functions whose hash values are close to those of disclosed vulnerable functions~\cite{tan2024similar}. 
We select these two tools for the following reasons: \ding{182} \textbf{High Efficiency}: They exhibit exceptional scanning speeds, enabling the scalable analysis of large-scale code bases. \ding{183} \textbf{Adjustable Coverage}: They allow us to maximize detection coverage by simply lowering the retrieval threshold parameter. \ding{184} \textbf{Targeted Scanning}: Crucially, they allow users to explicitly select the original vulnerabilities for scanning (thereby avoiding duplicate scanning), which is not supported by similar tools like VUDDY~\cite{vuddy}.
Finally, we take the union of their detection results as the outcome of the vulnerability detection phase to maximize the coverage of detected vulnerabilities, thereby significantly reducing the likelihood of overlooking potential security issues. 
It is worth noting that ReDeBug and the hash-based method are static detection tools, and they are pluggable.



\subsubsection{Vulnerability Confirmation}



This module is designed to further verify whether a detected vulnerability is indeed a real vulnerability rather than a false positive. 
The main challenge of vulnerability confirmation lies in how to obtain \textbf{sufficient information}. 
Most existing vulnerability detection approaches tend to focus on the analysis of the internal logic of a single function, while paying little attention to context information outside the function. 
To address this challenge, we leverage the comprehensive information of the original disclosed vulnerability to support agent-based decision-making, helping identify useful information including relevant functions and critical data structures. 
Subsequently, we design a set of tools, implemented using ripgrep~\cite{ripgrep} and Tree-sitter~\cite{treesitter}, to assist the agent in acquiring such information. 
There are four tools in total, which are used respectively to retrieve function and structure definitions from the original vulnerability repository and the target repository.

First, based on the analysis points of the original disclosed vulnerability stored in the VKB, the \portingagent adapts the analysis points to the new context. 
Then, the \analyzingagent uses these adapted analysis points to sequentially examine the suspicious functions in the target repository, ultimately determining whether the issue constitutes a true vulnerability. 
Meanwhile, with the help of these tools, we define a strategy to support the analysis: If the implementation of a function in the target repository is identical to that in the original vulnerability repository, then this function does not affect the occurrence of the vulnerability. 

\begin{figure}[htbp]
\centering
  \includegraphics[width=1\linewidth]{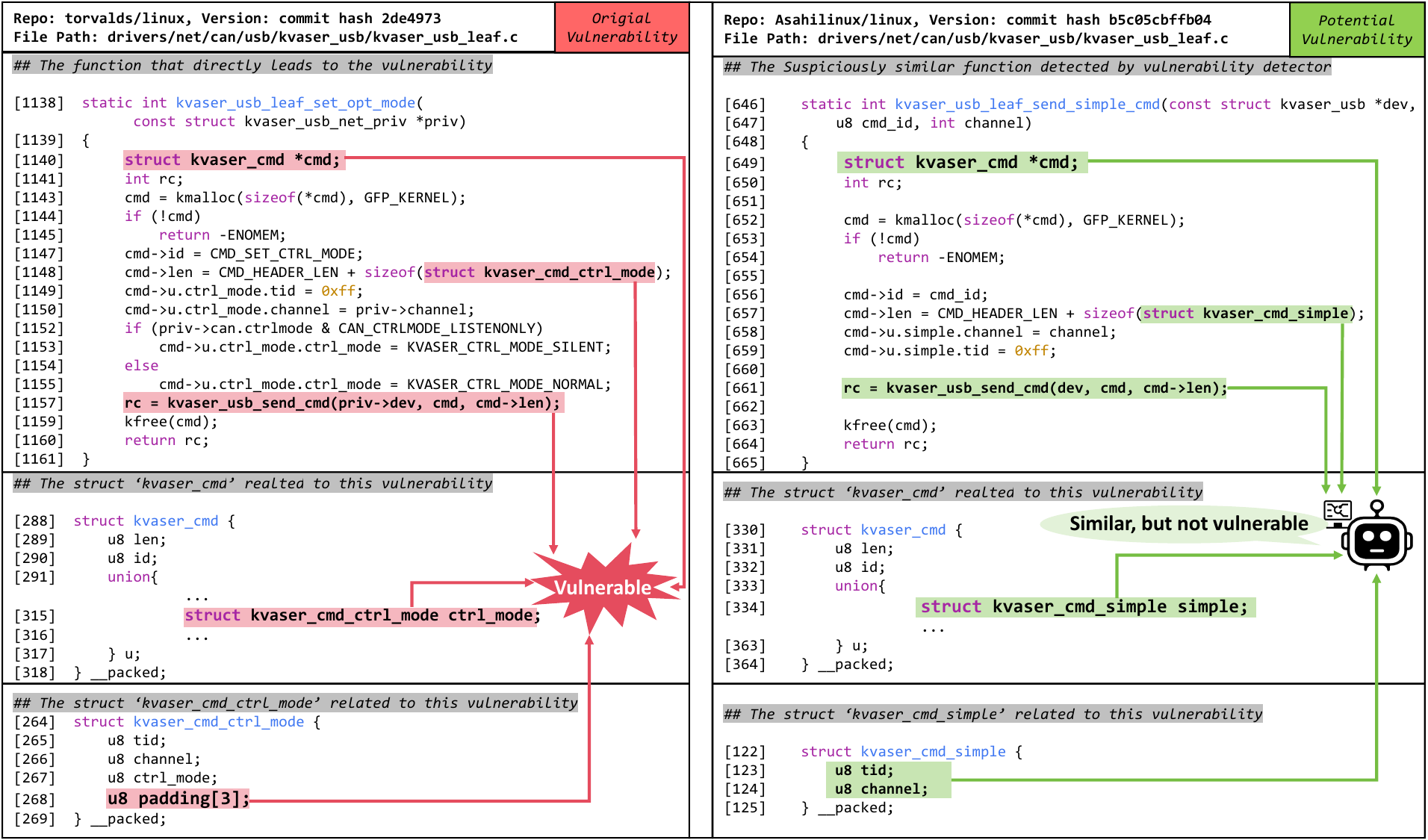}
  \caption{Using the vulnerability confirmation component to resolve a false positive in a function similar to CVE-2019-19947.}
  \label{fig:analysis-eg}
  \vspace{-0.3cm}
\end{figure}


To illustrate how the confirmation component works, we use the disclosed vulnerability CVE-2019-19947~\cite{CVE-2019-19947} from the torvalds/linux repository~\cite{torvalds} as an illustrative example. 
This vulnerability is fixed in commit da2311a~\cite{linux-commit-da2311a}. 
The root cause of the vulnerability is as follows: as shown on the left side of Figure~\ref{fig:analysis-eg}, within the file \path{drivers/net/can/usb/kvaser_usb/kvaser_usb_leaf.c}, the function \texttt{kvaser\_usb\_leaf\_set\_opt\_mode} defines a struct-type pointer \texttt{cmd} (line [1140]) referencing the structure \texttt{kvaser\_cmd} (lines [288]–[318]). 
In practice, the valid part of \texttt{cmd} corresponds only to the \texttt{kvaser\_cmd\_ctrl\_mode} structure (lines [264]–[269]). 
However, the \texttt{padding} fields are not initialized within the structure before passing \texttt{cmd} out (line [1157]). Moreover, since \texttt{kmalloc} does not zero-initialize memory, this results in the use of partially uninitialized variables, which may lead to memory leakage.

As shown on the right side of Figure~\ref{fig:analysis-eg}, using the vulnerability detection component, we identify a potentially vulnerable similar function \texttt{kvaser\_usb\_leaf\_send\_simple\_cmd} in the Asahilinux/linux repository~\cite{asahilinux} (version: commit hash b5c05cbffb04) at the same file path. 
In this function, a struct-type pointer \texttt{cmd} is also created (line [649]) for the \texttt{kvaser\_cmd} structure (lines [330]–[364]), but this function is not vulnerable. 
The \analyzingagent first identifies that the actual valid part of \texttt{cmd} corresponds to the \texttt{kvaser\_cmd\_simple} structure. 
By leveraging the pre-equipped tool capable of retrieving the concrete definition of a structure based on its name, it further obtains the definition of this structure (lines [122]–[125]) and finds that all fields in the structure are correctly initialized in similar function (lines [658]–[659]). 
Therefore, \appname successfully recognizes that this case is a false positive.

\subsection{Vulnerability Repair and Patch Validation}
\label{sec:repairandvalidation}
\subsubsection{Vulnerability Repair}

\begin{figure}[htbp]
  \includegraphics[width=1\linewidth]{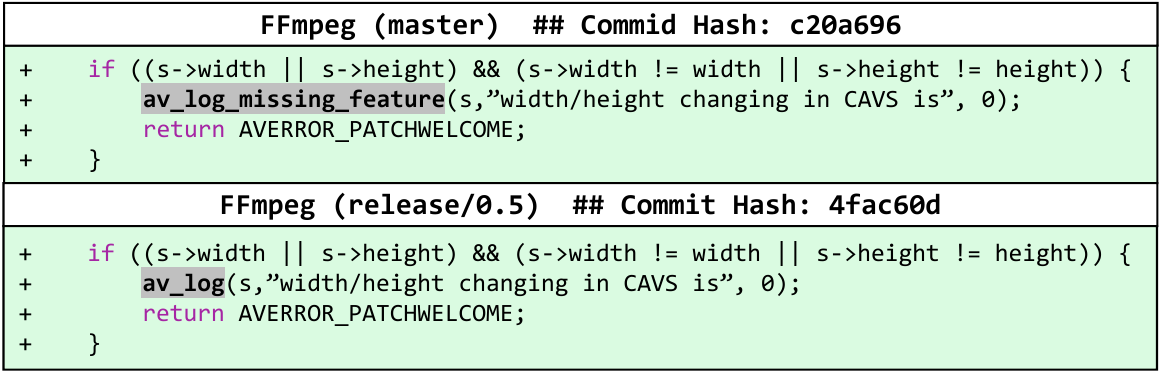}
  \caption{A patch porting example from FFmpeg-master to FFmpeg-release/0.5.}
  \label{fig:fixexample}
\vspace{-0.3cm}
\end{figure}

Similar to the confirmation phase, the fixing process also requires sufficient contextual information and faces a new challenge: \textbf{cross-repository consistency}. 
For example, functions implementing a certain feature may vary across different repositories. 
Figure~\ref{fig:fixexample} shows part of a patch pair, illustrating a patch transplanted from the FFmpeg master branch (upper part, commit hash: c20a696~\cite{ffmpeg-commit-c20a696}) to the FFmpeg release/0.5 branch (lower part, commit hash: 4fac60d~\cite{ffmpeg-commit-4fac60d}). 
The function \texttt{av\_log\_missing\_feature} exists only in the master branch, and in the release/0.5 branch, it is replaced with \texttt{av\_log} in the developer patch.
Inspired by this, we introduce the \consistencyagent. 
Using dedicated analysis tools, it examines whether the functions involved in the historical patch are consistent between the original and target repositories (e.g., function names and parameters). 
Then, the \consistencyagent generates a consistency report, laying the foundations for the subsequent repair process.
It is worth noting that this also serves as evidence for the application of repository-level context acquisition within our approach.

Based on (1) the analysis process of the current vulnerability, (2) the vulnerable function and the patch of the original vulnerable function, and (3) the consistency analysis report, the \fixingagent generates a patch for the current vulnerability.
When the consistency analysis report determines that a needed function does not exist in the target repository and the original patch, the \fixingagent searches for a semantically similar substitute function within the target repository by using the context tools shown in Figure ~\ref{fig:method}. 

\subsubsection{Patch Validation}
Patch validation focuses on determining whether it successfully addresses the identified vulnerability. 
We first reapply the original vulnerability detector to assess whether the patched code still contains the vulnerability. 
If the detector fails to identify any vulnerable code snippet in the patched function, \appname considers the vulnerability successfully fixed. 
Otherwise, if the detector reports a vulnerable code snippet, the \validationagent is employed to further verify whether the detection result is a false positive, i.e., whether the vulnerability has indeed been correctly fixed. 
If the \validationagent also determines that the vulnerability still exists, it provides feedback to the \fixingagent, which will then attempt to repair the code again. 
The prompt design and reasoning process of the \validationagent are almost identical to those of the \analyzingagent, except that the target functions for analysis are the patched functions rather than the original vulnerable ones. 
In our setting, we cap the validation-feedback loop to a single iteration for efficiency.


\section{Experiment Design}
\label{sec:experiment_design}

This section introduces the research questions, describes the data collection procedure, and presents the experimental settings, including baselines, evaluation metrics, and implementation details.

\subsection{Research Questions}
In this paper, we aim to answer the following research questions:

\noindent\textbf{RQ1: How effective is \appname compared to state-of-the-art methods?} 

\noindent\textbf{RQ2: What is the significance of the VKB component and the vulnerability confirmation component for \appname?} 


\subsection{Dataset}
\label{sec:dataset}

Although there is currently no directly available dataset for end-to-end vulnerability assessment, we observe that existing CVE patch porting cases meet our needs. 
They typically contain the following information:
(1) CVE ID, (2) source commit ($\mathit{commit_{s}}$), (3) target commit ($\mathit{commit_{t}}$), (4) source repository ($\mathit{repo_{s}}$), (5) target repository ($\mathit{repo_{t}}$), (6) pre-patch function in the source repository ($\mathit{f_{opre}}$), (7) post-patch function in the source repository ($\mathit{f_{opost}}$), (8) pre-patch function in the target repository ($\mathit{f_{tpre}}$), and (9) post-patch function in the target repository ($\mathit{f_{tpost}}$).
We use the version of $\mathit{repo_{t}}$ immediately preceding $\mathit{commit_{t}}$, which contains the vulnerable function $\mathit{f_{tpre}}$, as the input codebase.
Ideally, \appname should be able to detect and correctly repair $\mathit{f_{tpre}}$, while not falsely identifying vulnerabilities elsewhere. 
Thus, datasets from patch porting scenarios are well-suited for evaluating \appname. 
Notably, during data processing, we remove commits whose file paths contained keywords such as \textit{test} or \textit{version}, as these are likely unrelated to vulnerabilities' root causes.

We construct our evaluation benchmark by integrating and refining two source datasets: the Mystique dataset from Wu et al.'s work ~\cite{mystique} and the PPatHF dataset from Pan et al.'s work~\cite{pan2024automating}.
The rationale for choosing these two sources is twofold. 
First, they represent the latest and most comprehensive benchmarks in the field of patch porting, providing high-quality ground truth from real-world software repositories. 
Second, they offer complementary data distributions: Mystique provides a rich set of security-critical cases specifically focused on vulnerability patches across different branches, while PPatHF contributes extensive evolutionary data from long-term hard fork histories (i.e., Vim to Neovim). 
By integrating these two source benchmarks, we establish a diverse and representative foundation. 

\textbf{Step 1: Refinement of Mystique Dataset}. 
We first examine the evaluation dataset from Mystique. 
Upon inspection, we observe that for 603 out of the 694 CVEs, the patches in the source and target repositories are identical (i.e., possessing the same modified lines and surrounding context, resulting in a perfect match in \texttt{git diff}). 
We categorize these as trivial porting cases that do not require complex semantic adaptation, and thus offer limited value for evaluating the effectiveness of recurring vulnerability management in diverged codebases. 
Consequently, we filter out these identical cases and retain only those where the git diff outputs differ while the number of modified files remains consistent. 
This constraint ensures a determinate file-level correspondence between the source and target patches, facilitating accurate comparison. 
Additionally, we remove examples with inaccessible commit URLs. 
This process yields 48 CVEs, corresponding to 103 function-level patch pairs.

\textbf{Step 2: Extension with PPatHF Methodology.} 
To enhance the dataset's coverage of hard forks, we adopt the data collection methodology used in Pan et al.'s work~\cite{pan2024automating}. 
We extend their work by collecting all CVE patch transplantation records from Vim to Neovim up to April 21, 2025. 
Applying the same non-identical criterion as in Step 1 to exclude trivial ports, we obtain 59 CVEs, corresponding to 74 function-level patch migration pairs.

\begin{table}
  \centering
  \caption{Collected cases of patch porting across repositories. ``Intra-project'' denotes patch porting between different branches of the same repository, while ``Inter-project'' represents patch porting across forked projects.}
  \begin{tabular}{clc}
    \toprule
    \textbf{Relation Type} & \textbf{Repository Pair} & \textbf{\# Stars} \\
    \midrule
    \multirow{12}{*}{\shortstack{Intra-project\\(Different Branches)}}  
          & apache/httpd~\cite{httpd} & 3.7k  \\
          & FFmpeg/FFmpeg~\cite{ffmpeg} & 51.9k \\
          & FreeRDP/FreeRDP~\cite{freerdp} & 12k   \\
          & krb5/krb5~\cite{krb5} & 562   \\
          & xen-project/xen~\cite{xen} & 716   \\
          & OISF/suricata~\cite{suricata} & 5.5k  \\
          & openssl/openssl~\cite{openssl} & 28.3k \\
          & OpenVPN/openvpn~\cite{openvpn} & 12.3k \\
          & python/cpython~\cite{cpython} & 68.3k \\
          & qemu/qemu~\cite{qemu} & 11.8k \\
          & sqlite/sqlite~\cite{sqlite} & 8.2k  \\
    \midrule
    \multirow{2}{*}{\shortstack{Inter-project\\(Fork-based)}} 
          & vim/vim~\cite{vim} & 38.7k \\
          & neovim/neovim~\cite{neovim} & 91.8k \\
    \bottomrule
  \end{tabular}
  \label{tab:data}%
\vspace{-0.3cm}
\end{table}%


\textbf{Step 3: Further Filtering.} 
Subsequently, the datasets collected in Steps 1 and 2 are integrated, and the following filtering criteria are applied.
(1) Since our framework targets repository-specific vulnerability management, commits flagged as orphaned or not belonging to any branch are treated as invalid data and removed (see example in ~\cite{github-linux}). 
(2) We exclude samples related to the repo Wireshark~\cite{wireshark} primarily due to its highly template-driven architecture. 
Thousands of protocol dissector functions exhibit nearly identical structural patterns, which leads to severe distribution skew in preliminary detections based on code similarity~\cite{wireshark_docs}. 
This skew ultimately undermines the interpretability of the results.

As a result, we obtain 78 CVEs (i.e., 78 patch migration examples), corresponding to 114 function-level patch pairs. 
The dataset involves 13 widely used open-source repositories, as shown in Table~\ref{tab:data}. 
As C remains the predominant language in contemporary vulnerability research, our analysis is restricted to vulnerabilities residing in the C code.

\subsection{Baselines}

\noindent\textbf{\ding{172} FVF}~\cite{tan2024similar} identifies and filters SBP instances in vulnerability detection by analyzing code change histories. 
We compare it with our approach in terms of vulnerability confirmation capability. 
\noindent\textbf{\ding{173} PPatHF}~\cite{pan2024automating} is an LLM-based approach that automatically ports patches for hard forks on a function-wise basis. 
We compare it with our approach in terms of vulnerability repair capability.
\noindent\textbf{\ding{174} Mystique}~\cite{mystique} is a novel method to achieve automated vulnerability patch porting with semantic and syntactic enhanced LLM. 
We compare it with our approach in terms of vulnerability repair capability.
\noindent\textbf{\ding{175} GPT-4o}~\cite{gpt4o} is one of the most advanced closed-source LLMs. 
We also investigate the use of the off-the-shelf LLM for this task by including GPT-4o as a baseline. 
Moreover, we implement our approach using the GPT-4o as the underlying LLM. 
Thus, comparing \appname with the GPT-4o, we can better understand the shortcomings of directly applying the LLM in the vulnerability management task and the effectiveness of our design. 
The prompt used with GPT-4o is as similar as possible to those used in our own method.

\subsection{Evaluation Metrics}
\label{metrics}

Since our framework performs end-to-end vulnerability management, we assess its performance in two distinct phases: Vulnerability Detection and Vulnerability Repair. 

\textbf{Vulnerability Detection Metrics.} 
The detection performance is evaluated using standard metrics derived from the confusion matrix. 
Specifically, we focus on the True Positive (TP), False Positive (FP), and False Negative (FN). 
We define the metrics as follows:
\begin{itemize}
\item \textbf{TP:} 
The number of actual vulnerable functions correctly identified and confirmed. 
\item \textbf{FP:} 
The number of non-vulnerable functions incorrectly reported as vulnerabilities. 
\item \textbf{FN:} 
The number of actual vulnerable functions that the approach fails to identify. 
\end{itemize} 
True Negatives (TN) are excluded from our evaluation, as defining non-vulnerable negatives in a large-scale repository scanning scenario is practically infeasible and statistically meaningless. 
Based on these metrics, we further calculate \textbf{Precision ($\frac{TP}{TP+FP}$)}, \textbf{Recall ($\frac{TP}{TP+FN}$)}, and \textbf{F1-score ($\frac{2 \times Precision \times Recall}{Precision + Recall}$)} to provide a comprehensive view of the detection capabilities. 
High precision indicates a low false alarm rate, while high recall indicates that the system misses fewer vulnerabilities. 

\textbf{Vulnerability Repair Metrics.}
For the repair phase (which encompasses both patch generation and validation), we evaluate the quality of the fixes using the following two metrics: 
\begin{itemize}
\item \textbf{TPC/TP:}
TPC refers to the number of TP samples that are correctly repaired by an approach. 
Following Wu et al.'s work~\cite{mystique}, a patch is considered successful if it is semantically equivalent to the developer's patch. 
This metric reflects the true repair ability of an approach. 
\item \textbf{Accuracy} ($\frac{TPC}{TP + FP}$)\textbf{:} 
We define Repair Accuracy as the ratio of successfully fixed vulnerabilities to the total number of vulnerabilities identified and attempted by the method. 
The denominator includes all reported issues (potentially including false positives), while the numerator counts only the verified successful fixes. 
Therefore, this metric also serves as the definitive indicator of the framework's \textbf{end-to-end effectiveness} in recurring vulnerability management.


\end{itemize}







\subsection{Implementation Details}
The implementation details of \appname are as follows. 
We implement \appname using Python 3.12, based on the Langgraph~\cite{langgraph} framework. 
All experiments are conducted on Ubuntu 20.04.4 LTS. 
All software dependencies, including compilers, Python, and relevant libraries, are installed in a controlled environment to ensure reproducibility of results. 
For the LLM used in \appname, we mainly employ GPT-4o. 
Specifically, we set the LLM temperature to 0.

\section{Experiment Results}
\label{sec:experiment_result}

\subsection{RQ1. The Effectiveness of \appname}
\label{sec:rq1}

\begin{table}[htbp]
  \centering
  \caption{The Final Vulnerability Detection Results of \appname and baselines}
  \resizebox{1.0\linewidth}{!}{
    \begin{tabular}{lcccccc}
    \toprule
    \textbf{Approach} & \textbf{TP} & \textbf{FP} & \textbf{FN} & \textbf{Precision} & \textbf{Recall} & \textbf{F1-score} \\
    \midrule
    \textbf{MAVM} & 68    & \textbf{21} & 46    & \textbf{76.4\%} & 59.6\% & \textbf{67.0\%} \\
    ReDeBug + Hash + FVF & 78    & 219   & 36    & 26.3\% & 68.4\% & 38.0\% \\
    ReDeBug + Hash + GPT-4o & \textbf{81} & 112   & \textbf{33} & 42.0\% & \textbf{71.1\%} & 52.8\% \\
    \bottomrule
    \end{tabular}%
    }
  \label{tab:rq1-1}%
\end{table}%


\begin{table}[htbp]
  \centering
  \caption{The Final Vulnerability Repair Results of \appname and baselines}
  \resizebox{1.0\linewidth}{!}{
    \begin{tabular}{lcc}
    \toprule
    \textbf{Approach} & \textbf{TPC/TP} & \textbf{Accuracy} \\
    \midrule
    \textbf{MAVM} & \textbf{51/68 (75.0\%)} & \textbf{57.3\%} \\
    ReDeBug + Hash + FVF + PPatHF & 47/78 (60.3\%) & 15.8\% \\
    ReDeBug + Hash + FVF + GPT-4o & 36/78 (46.2\%) & 12.1\% \\
    ReDeBug + Hash + FVF + Mystique & 46/78 (59.0\%) & 15.5\% \\
    ReDeBug + Hash + GPT-4o + PPatHF & 49/81 (60.5\%) & 25.4\% \\
    ReDeBug + Hash + GPT-4o + GPT-4o & 37/81 (45.7\%) & 19.2\% \\
    ReDeBug + Hash + GPT-4o + Mystique & 48/81 (59.3\%) & 24.9\% \\
    \bottomrule
    \end{tabular}%
    }
  \label{tab:rq1-2}%
\end{table}%

\begin{table*}[htbp]
  \centering
  \caption{The performance of various variants of \appname}
    \begin{tabular}{lcccccccc}
    \toprule
    \multirow{2}[4]{*}{\textbf{Approach}} & \multicolumn{6}{c}{\textbf{Detection Phase}}  & \multicolumn{2}{c}{\textbf{Repair Phase}} \\
\cmidrule(lr){2-7} \cmidrule(lr){8-9}           & \textbf{TP} & \textbf{FP} & \textbf{FN} & \textbf{Precision} & \textbf{Recall} & \textbf{F1-score} & \textbf{TPC/TP} & \textbf{Accuracy} \\
    \midrule
    \textbf{MAVM} & 68    & \textbf{21} & 46    & \textbf{76.4\%} & 59.6\% & \textbf{67.0\%} & \textbf{51/68 (75.0\%)} & \textbf{57.3\%} \\
    MAVM-c & \textbf{81} & 223   & \textbf{33} & 26.6\% & \textbf{71.1\%} & 38.8\% & 54/81 (66.7\%) & 17.8\% \\
    MAVM-v & 69    & 32    & 45    & 68.3\% & 60.5\% & 64.2\% & 50/69 (72.5\%) & 49.5\% \\
    \bottomrule
    \end{tabular}%
  \label{tab:rq2}%
\end{table*}%

\noindent\textbf{Setup}. This section aims to evaluate the effectiveness of \appname in end-to-end vulnerability management, using GPT-4o as the underlying LLM for \appname. 
Since no existing approach can achieve complete end-to-end vulnerability management, we construct baseline methods by combining state-of-the-art approaches from each phase. 
As the detection component of \appname is fully pluggable, we standardize the detection setup for all methods by using the same static detection tools (ReDeBug~\cite{redebug} and hash-based method~\cite{tan2024similar}) with identical parameter settings for all methods. 
This uniform configuration ensures a fair comparison, allowing us to attribute any performance gains solely to our core innovations in the multi-agent framework and the VKB.
In the confirmation phase, we employ FVF and GPT-4o as baselines. 
In the repair phase, we employ PPatHF, Mystique, and GPT-4o as baselines. 
These methods are combined in pairs to produce six hybrid methods, as shown in the column `Approach` of Table \ref{tab:rq1-2}.
For GPT-4o used in the confirmation and repair phases, the input mainly includes the original vulnerability function, the original patch content, and the new vulnerability function. 
We configure Mystique with GPT-4o as its base LLM to maintain consistency with our approach. 
This choice is necessitated by the fact that the original Mystique study~\cite{mystique} involves multiple specialized fine-tuned models, which makes it impractical to select a single representative baseline without introducing bias. 
In contrast, for PPatHF, we strictly follow its original configuration by utilizing the specific fine-tuned version of StarCoder as described in~\cite{pan2024automating}.

\noindent\textbf{Results and Analyses}. 
The experimental results are presented in Table~\ref{tab:rq1-1} and Table~\ref{tab:rq1-2}. 
Table~\ref{tab:rq1-1} reports the results of the vulnerability detection (comprising both initial detection and subsequent confirmation), while Table~\ref{tab:rq1-2} focuses on the vulnerability repair results. 
Compared to the hybrid baselines, \appname achieves the highest repair accuracy (57.3\%), highlighting its superior effectiveness in end-to-end management.

\noindent\textbf{Detection Results}. 
While all methods start with the same static tools and parameters, the performance diverges during the confirmation phase. 
As shown in Table~\ref{tab:rq1-1}, the FVF-based baseline fails to filter FPs effectively, resulting in 219 FPs and a low precision of 26.3\%. 
This is because FVF is limited to specific patch-migration patterns and cannot handle complex cases. 
While GPT-4o improves precision to 42.0\%, it still produces 112 FPs.
In contrast, \appname achieves the best precision and F1-score with only 21 FPs. 
This superior filtering stems from the VKB component, which provides deep insights into the root causes and trigger chains of historical vulnerabilities. 
Furthermore, the Analyzing Agent retrieves real-time context using specialized tools to distinguish true vulnerabilities from safe code. 
Although \appname's recall (59.6\%) is lower than the approach using GPT-4o as the confirmation phase (71.1\%) due to a more rigorous confirmation strategy, its high-precision output provides a much cleaner foundation for the repair phase. 
Both FVF and GPT-4o struggle to distinguish real vulnerabilities from the initial detection results. 
Consequently, despite their higher recall, they generate a substantial number of FPs.

\noindent\textbf{Repair Results}. 
Since \appname is a fully integrated end-to-end framework, it cannot be decomposed into isolated patching tools for direct comparison with individual patch-porting methods. 
Therefore, we construct baselines by combining state-of-the-art tools to represent current capabilities. 
As shown in Table~\ref{tab:rq1-2}, \appname successfully fixes 51 vulnerabilities. 
In the TPC/TP metric, \appname reach 75.0\%, outperforming baselines by 14.5\%–29.3\%, which demonstrates the repair effectiveness of \appname. 
Regarding the accuracy metric, \appname achieves 57.3\%, outperforming the baselines by a significant margin of 31.9\%–45.2\%. 
As this metric reflects the end-to-end success rate (see Section~\ref{metrics} for details), these results demonstrate the superior effectiveness of \appname in the end-to-end vulnerability management.
The performance gap in repair is driven by three factors. 
First, the baselines suffer from the detection stage's limitations, as they attempt to repair numerous FPs where no valid fix is possible. 
Second, whereas baselines depend exclusively on the original patch, the Fixing Agent in \appname incorporates the analytical insights from \analyzingagent to gain a deeper understanding of the vulnerability and utilizes specialized tools to retrieve extensive repository-level context.
Furthermore, \consistencyagent verifies the consistency of the functions required in the patch by confirming their existence and the alignment of their parameters. 
These consistency results serve as essential contextual information, which provides a robust foundation for the repair.
These results prove that the context reasoning and VKB-derived insights significantly enhance the effectiveness of recurring vulnerability detection and repair.

\begin{tcolorbox}[colback=gray!5!white, colframe=black, boxrule=0.5pt, left=2pt, right=2pt, top=2pt, bottom=2pt]
\textbf{Answer to RQ1:} Overall, \appname demonstrates superior performance compared to the baselines across all phases, indicating its high efficiency. Notably, its enhanced ability to filter false positives and its superior repair accuracy represent a significant step toward making automated vulnerability management more practical for real-world security auditing.
\end{tcolorbox}

\subsection{RQ2. The Key Designs of \appname}
\label{sec:rq2}

\noindent\textbf{Setup}. 
This section investigates the importance of the VKB component and the vulnerability confirmation component in \appname through ablation experiments. 
We use GPT-4o as the base LLM for \appname and other variants. 
We denote the variant of \appname without the VKB component as \appnamev, and the variant without the confirmation component as \appnamec.

\noindent\textbf{Result and Analyses.} 
As shown in Table~\ref{tab:rq2}, \appname achieves the best performance in terms of Precision, F1-score, TPC/TP, and Repair Accuracy. 
Specifically, in the end-to-end task, \appname exceeds \appnamev and \appnamec in repair accuracy by 7.8\% and 39.5\%, respectively.

\noindent\textbf{The Role of Vulnerability Confirmation.} 
The most significant performance drop occurs when the confirmation component is removed (\appnamec). 
Without this module, the number of FPs surges from 21 to 223, causing precision to plummet from 76.4\% to 26.6\%. 
While \appnamec achieves the highest recall (71.1\%) because it skips the rigorous confirmation filter, it introduces a massive influx of noise. 
This demonstrates that the confirmation component acts as a critical quality gate, ensuring that the framework focuses its repair efforts on high-confidence vulnerabilities.
Furthermore, the decrease in TPC/TP (8.3\%) for \appnamec indicates that the analysis process within the confirmation phase is important, as it provides the \fixingagent with critical insights necessary to generate correct patches.
Ultimately, the overall repair accuracy of \appnamec falls to only 17.8\%, which further validates that our vulnerability confirmation phase is indispensable for end-to-end recurring vulnerability management.

\noindent\textbf{The Role of VKB.} 
The performance decline observed in \appnamev indicates that constructing the VKB significantly enhances the framework's capability to analyze vulnerabilities and suppress false detections. 
Specifically, removing the VKB component leads to an increase in FPs from 21 to 32, which causes the precision to drop from 76.4\% to 68.3\%. 
This suggests that the specialized knowledge within the VKB assists \appname in accurately distinguishing true vulnerabilities from non-vulnerable code. 
Furthermore, \appnamev exhibits a 2.5\% decrease in TPC/TP, which confirms that the absence of VKB-derived insights weakens the framework's repairing capacity. 
Ultimately, the overall repair accuracy of \appnamev falls to 49.5\%, which further validates that the VKB is indispensable for end-to-end recurring vulnerability management.

\begin{tcolorbox}[colback=gray!5!white, colframe=black, boxrule=0.5pt, left=2pt, right=2pt, top=2pt, bottom=2pt]
\textbf{Answer to RQ2:} The VKB component lays the foundation for subsequent vulnerability management, which facilitates precise vulnerability analysis and suppresses false detections.
The vulnerability confirmation module effectively analyzes detected vulnerabilities to reduce false positives, and its analysis process also facilitates successful repairs. 
\end{tcolorbox}

\section{Discussion}
\label{sec:discussion}

\subsection{Case Study}
To illustrate the effectiveness of \appname, we select the real-world vulnerability identified as CVE-2023-4752~\cite{CVE-2023-4752} from the vim/vim~\cite{vim} as the original vulnerability, and use the neovim/neovim~\cite{neovim} (version: commit hash a589156) as the target repository to perform the vulnerability management.

\textbf{Vulnerability Knowledge Collection and Construction}.
CVE-2023-4752 involves unsafe buffer operations where memory is freed without prior validity checks, leading to a use-after-free condition. 
The corresponding patch in Vim introduces explicit buffer validation logic~\cite{casestudy}. 
\appname automatically extracts such key historical vulnerability features from the vulnerability's description and patch, providing essential context for subsequent reasoning.

\textbf{Context-Aware Vulnerability Detection and Verification}.
Through ReDeBug and hash-based detection, the functions \texttt{set\_init\_fenc\_default} and \texttt{ins\_compl\_get\_exp} are both flagged as potential vulnerable functions.
In fact, \texttt{set\_init\_fenc\_default} is not vulnerable, and \appname successfully identifies this.
First, the \portingagent migrates the analysis points of the historical vulnerability to the current context, which aims to verify whether invalid buffer accesses exist in the function. 
Then, the \analyzingagent, guided by these points, searches for relevant contextual clues using context tools and finds that this function only deals with default encoding initialization and memory allocation, which is unrelated to the invalid buffer accesses.
Thus, the function is correctly determined to be non-vulnerable.
Following a similar analysis process, MAVM successfully identifies \texttt{ins\_compl\_get\_exp} as a recurring vulnerability.
These cases demonstrate that \appname can effectively leverage historical knowledge and agents with useful tools to identify vulnerabilities in target repositories with high precision.

\textbf{Knowledge-Guided Repair and Validation}.
For the confirmed vulnerable function \texttt{ins\_compl\_get\_exp}, the \consistencyagent first examines the \texttt{buf\_valid} function referenced in the original vulnerability patch and finds inconsistencies (e.g., the return type differs: \texttt{int} in the original repository versus \texttt{bool} in the target repository). 
Subsequently, the \fixingagent, guided by the \analyzingagent's analysis process, combines insights from the original patch and the consistency-check result to introduce a validation mechanism for the current function, successfully fixing the vulnerability.
After patch generation, the hash-based detector still flags this function as vulnerable because its code structure remains similar to the original vulnerable pattern.
However, the \validationagent, following the same analysis process as the \analyzingagent, confirms that the patch correctly eliminates the vulnerability. 
This further demonstrates that \appname achieves precise vulnerability detection, and that consistency checking is an indispensable step during vulnerability repair. 
Finally, the generated patch is automatically applied to the original repository, thereby achieving a closed-loop, end-to-end vulnerability management process.

This case highlights why the architecture of \appname is highly efficient:
(1) \textbf{Historical Knowledge}. 
Before performing vulnerability detection, \appname conducts an in-depth analysis of the original vulnerability, obtaining crucial information such as its category and patching strategy, which effectively guides subsequent phases of vulnerability management. 
(2) \textbf{Contextual Intelligence}. 
Both the \analyzingagent and the \consistencyagent leverage context-aware tools to perform repository-level reasoning. 
(3) \textbf{Rigorous Pipeline}. 
\appname establishes a tightly integrated workflow: it not only detects and repairs vulnerabilities but also performs confirmation and post-repair validation. 
Moreover, the vulnerability detector and the \validationagent serve as dual safeguards to ensure patch correctness in the validation phase.

\subsection{Threat to Validity}
\noindent\textbf{Internal Validity}. 
The main threat to internal validity lies in the randomness of LLM outputs. 
During the experiments, we mitigate this threat by setting the API temperature parameter to 0 and providing pre-selected references in the prompts to minimize the model's degree of freedom. 
Due to the cost limitations associated with invoking the API, the number of experimental runs in this study is limited. 
In the future, we plan to conduct repeated experiments to further investigate the impact of LLMs' randomness on our approach.

\noindent\textbf{External Validity}. 
The primary threat to external validity lies in the choice of dataset. 
While the architecture of \appname is theoretically language-agnostic, our current evaluation focuses exclusively on the C language. We plan to extend our experiments to include a broader range of programming languages in future work to further demonstrate the generalizability of the approach. 
Due to the high complexity of dataset collection, preparation, and experimentation for the full vulnerability management process, our dataset size is relatively small. 
In the future, we will expand the dataset size and expand the dataset to include repositories with a wider variety of source code types.

\section{Conclusion and Future Work}
\label{sec:conclusion}


In this paper, we propose \appname, the first end-to-end vulnerability management approach aimed at recurring vulnerabilities. 
Based on agents, MAVM effectively simulates real-world security workflows.
In \appname, we construct the VKB from disclosed historical vulnerabilities, which addresses the insufficient use of original vulnerability information in previous approaches, while also mitigating the lack of domain-specific knowledge in LLMs. 
By pre-designing tools capable of retrieving repository contextual information, agents can autonomously invoke these tools to acquire sufficient context, thereby resolving the issue of missing contextual information during vulnerability management tasks. 
We demonstrate the superiority of \appname's vulnerability management capabilities by comparing it against state-of-the-art approaches for each phase of the vulnerability management process.
For future work, we aim to extend vulnerability management from the current function-level to the CVE-level. 
Compared to the function-level, the CVE-level may allow for a more accurate grasp of the essence of a vulnerability, which would be highly beneficial for both its detection and repair.



\balance
\bibliographystyle{IEEEtran}
\bibliography{main}
\end{document}